\documentclass[final]{ws-procs9x6_arxiv}

\usepackage{bbm}

\newcommand{\Cx}{\mathbbm{C}}

\newcommand{\Nl}{\mathbbm{N}}

\newcommand{\Rl}{\mathbbm{R}}

\newcommand{\p}{\mathbbm{p}}
\newcommand{\q}{\mathbbm{q}}

\newcommand{\idty}{\mathbbm{1}}
\DeclareMathOperator{\id}{id}

 \DeclareMathOperator{\tr}{Tr}
\newcommand{\<}{\langle}
\renewcommand{\>}{\rangle}
\providecommand{\abs}[1]{\lvert#1\rvert}
\providecommand{\norm}[1]{\lVert#1\rVert}
\DeclareMathOperator*{\loplus}{\mbox{\Large\mbox{$\oplus$}}}

\renewcommand{\b}[1]{\mathbf{#1}}
\renewcommand{\c}[1]{\mathcal{#1}}
\newcommand{\g}[1]{\mathfrak{#1}}
\newcommand{\s}[1]{\mathsf{#1}}
\renewcommand{\r}[1]{\mathrm{#1}}
\newcommand{\bit}[1]{\boldsymbol{#1}}

\newcommand{\bra    }{\langle}
\newcommand{\ket    }{\rangle}

\begin{document}

\title{QUANTUM PROCESSES\footnote{to appear in the proceedings of the 46th Karpacz Winter School of Theoretical Physics "Quantum Dynamics and Information: Theory and Experiment"}}

\author{M. FANNES$^1$ and J. WOUTERS$^{2}$}

\address{Instituut voor Theoretische Fysica, K.U.Leuven, \\
3001-Heverlee, Belgium \\
$^1$E-mail: mark.fannes@fys.kuleuven.be \\
$^2$E-mail: jeroen.wouters@fys.kuleuven.be}

\begin{abstract}
A number of ideas and questions related to the construction of quantum processes are discussed. Quantum state extension, entanglement and asymptotic behaviour of the entropy are some of the issues explored.
These topics are studied in more detail for a class of quantum processes known as finitely correlated states. Several examples of such processes are presented, specifically a Free Fermionic model.
\end{abstract}


\bodymatter

\section{Introduction}
\label{s1}

The aim of this note is to present a number of ideas and questions related to the construction of quantum processes. For technical convenience we restrict ourselves to systems with a finite configuration space in a classical context or with $d$ accessible levels in the quantum.

A reversible dynamics of such an isolated system is rather boring: a finite classical configuration space only supports jumps in discrete time while the evolution of a finite level quantum systems is almost periodic in time
\begin{equation*}
A_t = \sum_\omega \r e^{i\omega t}\, \hat A(\omega).
\end{equation*}
Here, the summation runs over the Bohr frequencies of the system, i.e., the spacings between the energy levels of the Hamiltonian and the $\hat A(\omega)$ are the Fourier coefficients of the observable $A$.

The evolution of systems weakly perturbed by an environment can be reasonably described by a stochastic dynamics, even by a stochastic map $\Gamma$ if we only observe the system at regular time intervals. Such maps will typically turn pure states into mixed states, a clear signature of their randomizing character. In the classical context we deal with matrices of transition probabilities and in the quantum setting with completely positive trace preserving maps. Repeated action of the map on the states yields a dynamics $\{\Gamma^n \mid n \in \Nl\}$, Markovian in time. A generic $\Gamma$ has a unique invariant state $\rho_0$ into which any initial state is driven
\begin{equation*}
\Gamma^n (\text{initial state}) \ \stackrel{n \to \infty}{\longrightarrow}\  \rho_0.
\end{equation*}

Several figures of merit can be defined for this kind of evolution. Two well-known ones are
the spectral gap $\gamma$ of $\Gamma$ that controls the asymptotic rate of convergence towards the invariant state
\begin{equation*}
\norm{\Gamma^n(\rho) - \rho_0} \sim (1 - \gamma)^n,\enskip n \text{ large}
\end{equation*}
and the minimal output entropy
\begin{equation*}
\s H^{\text{min}}(\Gamma) := \min\Bigl( \bigl\{ \s H(\Gamma(\rho)) \,\bigm|\, \rho \text{ state} \bigr\} \Bigr).
\end{equation*}
Here $\s H$ denotes either the Shannon entropy of a probability vector or the von~Neumann entropy of a density matrix.

A stochastic process in discrete time is a different object: it specifies joint probabilities at different times. More precisely such a process is a state $\bit\omega$ on a half-chain $\Nl$. At each site of this chain sits a copy of our classical configuration space or of the quantum $d$-level system. We assume, moreover, that the chain is stationary: the state $\bit\omega$ is invariant under a right shift.

In the classical case the process models a source that is emitting every time unit a letter belonging to the alphabet $\{\epsilon_1,  \epsilon_2, \ldots, \epsilon_d\}$. The state $\bit\omega$ specifies the probability that the source emits a given word. Let $\bigl( \epsilon(0), \epsilon(1), \epsilon(2), \ldots \bigr)$ be a random letter string emitted by the source at times $t = 0, 1, 2, \ldots$ then
\begin{equation*}
\text{Prob}\Bigl( \epsilon(0) = \epsilon_0\ \&\ \epsilon(1) = \epsilon_1\ \&\ \ldots\ \&\  \epsilon(n) = \epsilon_n \Bigr) = \bit\omega(\epsilon_0, \epsilon_1, \ldots, \epsilon_n).
\end{equation*}

In the quantum context, the restriction of $\bit\omega$ to the first $n$ sites of the chain is a density matrix on $\otimes^n \c M_d$. This density matrix encodes all the statistical information that can be obtained by applying repeated measurements on a sequence of $n$ particles emitted by the source.

The amount of randomness in the process up to time $n$ is quantified by the entropy
\begin{equation*}
\s H_n = \s H\bigl( \bit\omega_{\{0,1,\ldots,n\}} \bigr).
\end{equation*} 
For stationary processes $\s H_n$ satisfies sub-additivity $\s H_{m+n} \le \s H_m + \s H_n$. This implies the existence of the average entropy
\begin{equation}
\s h := \lim_{n \to \infty} \tfrac{1}{n+1}\, \s H_n.
\label{av:ent}
\end{equation}
The quantity $\s h$ is also known as the dynamical entropy of the shift on the half-chain, it takes values in $[0, \log d]$. It is a relevant measure of the randomness of the source as, under suitable ergodicity conditions on $\bit\omega$, length-$n$ messages can be reliably encoded in a subspace of dimension $\exp(n \s h)$ instead of $d^n$.

Using strong sub-additivity, $\s H_{\ell + m + n} + \s H_m \le \s H_{\ell + m} + \s H_{m + n}$, more can be proven: the local entropy $n \mapsto \s H_n$ is increasing, while the entropy increment $n \mapsto \bigl( \s H_n - \s H_{n-1} \bigr)$ is decreasing. Both properties fail for general non shift-invariant quantum states. As a consequence
\begin{equation}
\s h = \lim_{n \to \infty}  \bigl( \s H_n - \s H_{n-1} \bigr).
\label{ent:pro}
\end{equation}
This means that $\s h$ is not only the compression rate of long messages but also the asymptotic entropy production of the source. The importance of~(\ref{ent:pro}) is that it can be used as a starting point for computing $\s h$. This happens for some Markov-like constructions where a simple transfer matrix-like construction generates the $n$-steps marginal of $\bit\omega$ from the $(n-1)$-th. E.g., Blackwell~\cite{bla} has described a procedure for computing the entropy of hidden Markov processes and we shall show that an analogous procedure applies to free Fermionic processes.

There are several routes that lead to classical Markov processes, like extending two-party states or generating the process in terms of a stochastic matrix. We remind here a number of problems and results that arise in this context with quantum processes. We also present a general scheme for generating quantum processes in terms of a quantum operation on a $d$-level system. Several examples are considered, in particular a free Fermionic version.

\section{Classical Markov Processes}
\label{s2}

The configuration space of a classical register with $d$ states is just a finite set $\Omega = \{ 1, 2, \ldots, d \}$. The states are length-$d$ probability vectors
\begin{equation*}
\bit\mu = \{ \mu(1), \mu(2), \ldots, \mu(d) \},\enskip \mu(\epsilon) \ge 0,\enskip \sum_\epsilon \mu(\epsilon) = 1.
\end{equation*}
The state space is a simplex and the extreme points are the Dirac measures that assign a probability 1 to a configuration. The Shannon entropy
\begin{equation*}
\s H(\bit\mu) := - \sum_\epsilon \mu(\epsilon) \log \mu(\epsilon)
\end{equation*}
quantifies the randomness in the state. It is easily seen that $\s H$ is a concave function on the state space:
\begin{equation*}
\s H(\lambda_1 \bit\mu_1 + \lambda_2 \bit\mu_2) \ge \lambda_1\, \s H(\bit\mu_1) + \lambda_2\, \s H(\bit\mu_2), \enskip \lambda_i \ge 0,\enskip \lambda_1 + \lambda_2 = 1.
\end{equation*} 

Restricting a state $\bit\mu_{12}$ on a composite system $\Omega_{12} = \Omega_1 \times \Omega_2$ to the subsystem $\Omega_1$ returns the first marginal of $\bit\mu_{12}$
\begin{equation*}
\bit\mu_1(\epsilon_1) := \sum_{\epsilon_2} \bit\mu_{12}(\epsilon_1,\epsilon_2).
\end{equation*}
The Shannon entropy behaves well with respect to restrictions:
\begin{itemize}
\item
monotonicity: $\s H(\bit\mu_1) \le \s H(\bit\mu_{12})$,
\item
sub-additivity: $\s H(\bit\mu_{12}) \le \s H(\bit\mu_1) + \s H(\bit\mu_2)$, and
\item
strong sub-additivity: $\s H(\bit\mu_{123}) + \s H(\bit\mu_2) \le \s H(\bit\mu_{12}) + \s H(\bit\mu_{23})$.
\end{itemize}

We can now consider the following state extension problem. Suppose that we are given two probability vectors $\bit\mu_{12}$ and $\bit\nu_{23}$ that agree on the middle system: $\bit\mu_2 = \bit\nu_2$. Can we find a joint extension for $\bit\mu_{12}$ and $\bit\nu_{23}$? More explicitly: can we find a state $\bit\xi_{123}$ on $\Omega_{123}$ that restricts to $\bit\mu_{12}$ on $\Omega_{12}$ and to $\bit\nu_{23}$ on $\Omega_{23}$? This is indeed possible and clearly the set of joint extensions $\bit\xi_{123}$ is convex and compact. We can therefore refine the question and ask for a joint extension of maximal entropy. A straightforward computation yields the answer:
\begin{equation}
\bit\rho_{123}(\epsilon_1, \epsilon_2, \epsilon_3) := \frac{\bit\mu_{12}(\epsilon_1, \epsilon_2)\, \bit\nu_{23}(\epsilon_2, \epsilon_3)}{\bit\mu_2(\epsilon_2)} = \frac{\bit\mu_{12}(\epsilon_1, \epsilon_2)\, \bit\nu_{23}(\epsilon_2, \epsilon_3)}{\bit\nu_2(\epsilon_2)}.
\label{mar:ext}
\end{equation}
Actually, $\bit\rho_{123}$ saturates the strong sub-additivity inequality:
\begin{equation*}
\s H(\bit\rho_{123}) + \s H(\bit\rho_2) = \s H(\bit\rho_{12}) + \s H(\bit\rho_{23}).
\end{equation*}
Unsurprisingly, there is a direct connection with thermal equilibrium states. If we introduce Hamiltonians
\begin{equation*}
\bit\mu_{12} = \r e^{-h_{12}},\enskip \bit\nu_{23} = \r e^{-h_{23}}, \enskip\text{and}\enskip \bit\mu_2 = \bit\nu_2 = \r e^{-h_2},
\end{equation*}
then
\begin{equation*}
\bit\rho_{123} = \r e^{-(h_{12} + h_{23} - h_2)}.
\end{equation*}

Let us start with a two-party probability vector $\bit\mu$ that is shift-invariant:
\begin{equation}
\sum_{\epsilon_2} \bit\mu(\epsilon, \epsilon_2) = \sum_{\epsilon_1} \bit\mu(\epsilon_1, \epsilon)\enskip \text{for all $\epsilon$}.
\label{inv}
\end{equation}
We can repeatedly apply the Markov extension procedure~(\ref{mar:ext}) to get a stationary process
\begin{equation}
\bit\omega(\epsilon_0, \epsilon_1, \ldots, \epsilon_n) = \frac{\bit\mu(\epsilon_0, \epsilon_1)\, \bit\mu(\epsilon_1, \epsilon_2)\, \cdots\, \bit\mu(\epsilon_{n-1}, \epsilon_n)}{\bit\mu(\epsilon_1)\, \bit\mu(\epsilon_2)\, \cdots\, \bit\mu(\epsilon_{n-1})}.
\label{mar:pro}
\end{equation}

Another procedure is to start with a $d \times d$ stochastic matrix $T$. The entry $T_{\epsilon_1 \epsilon_2}$ is the probability for jumping from state $\epsilon_1$ to $\epsilon_2$, therefore
\begin{equation}
T_{\epsilon_1 \epsilon_2} \ge 0 \enskip\text{and}\enskip \sum_{\epsilon_2} T_{\epsilon_1 \epsilon_2} = 1.
\label{tra:pro}
\end{equation}
The invariant state $\bit\mu$ is a row vector determined by $\bit\mu \, T = \bit\mu$. The Markov process is now obtained by putting
\begin{equation}
\bit\omega(\epsilon_0, \epsilon_1, \ldots, \epsilon_n) = \bit\mu(\epsilon_0)\, T_{\epsilon_0 \epsilon_1}\, \cdots\, T_{\epsilon_{n-1} \epsilon_n}.
\label{mar:pro2}
\end{equation}
Both constructions~(\ref{mar:pro}) and~(\ref{mar:pro2}) agree if we put
\begin{equation*}
T_{\epsilon_1 \epsilon_2} = \frac{\bit\mu(\epsilon_1, \epsilon_2)}{\bit\mu(\epsilon_1)}.
\end{equation*}

The rows of a stochastic matrix $T$ are probability vectors. The minimal output entropy of $T$ is simply
\begin{equation*}
\s H^{\text{min}}(T) = \text{smallest entropy of rows of $T$}
\end{equation*}
while the entropy of the process is a smooth version of this quantity
\begin{equation*}
\s h = \text{$\bit\mu$-average of entropies of rows of $T$}.
\end{equation*}

\section{Extending Quantum States}
When turning to quantum state extension the situation gets more complicated. Quantum states allow for more freedom, as they exhibit correlations that are not present in classical systems, but this imposes at the same time more stringent positivity conditions.

States on a full matrix algebra $\c M_d$ can be identified with density matrices: non-negative matrices with trace one. The convex set of density matrices is very unlike a simplex. A density matrix that is not an extreme point of the state space, i.e., that is not a one-dimensional projector, allows many decompositions in extreme states. In contrast with classical systems such a state can therefore not be seen as a well-defined ensemble op pure states. We need $d^2 -1$ real parameters to describe the state space of $\c M_d$ while $2d-2$ parameters suffice to label the pure states. This means that the boundary of the state space contains many flat subsets. Nevertheless the pure states form a very nice smooth manifold. The case of a single qubit is exceptional: its state space is affinely isomorphic to the Bloch ball by the standard parametrization
\begin{equation}
\rho = \tfrac{1}{2}\, (\idty + \b x \cdot \bit\sigma),\enskip \b x \in \Rl^3,\ \norm{\b x} \le 1.
\label{bloch}
\end{equation}
In this case, every point of the boundary is also an extreme point. For higher $d$, a smooth parametrization of the pure states does not define a boundary of a convex set.

For a composite system, restricting to a sub-system amounts to taking partial traces over remaining parties
\begin{equation*}
\rho_1 := \tr_2 \rho_{12}.
\end{equation*}
The entropy of a state with density matrix $\rho$ is given by the von Neumann entropy
\begin{equation*}
\s H(\rho) = - \tr \rho \log \rho.
\end{equation*}
However, already the most basic property of the Shannon entropy, monotonicity, does not carry over. Consider for example the maximally entangled two-qubit state $|\Phi^+ \ket \bra \Phi^+ |$ with $|\Phi^+\ket = (| 00 \ket + |11\ket)/\sqrt{2}$. Its entropy $\s H(|\Phi^+ \ket \bra \Phi^+ |)$ is zero as it is a pure state, while its restriction $\rho_1$ is the maximally mixed state, which has maximal entropy, so clearly $\s H(\rho_1) \nleq H(\rho_{12})$.

An important property that holds both for classical and quantum systems is that if the marginal $\rho_1$ of a bipartite state is pure then $\rho_{12} = \rho_1 \otimes \rho_2$. This is an important ingredient of the theory: it namely allows to isolate a system from the rest of the universe. At the same time it is also a severe constraint on quantum systems because there are plenty of pure states of a composite system. In particular the restriction of an entangled pure state can never be pure and we can therefore not separate a party of an entangled system from the outside world, which is more or less what goes wrong with the locality assumption in the EPR paradox.

Factorisation of extensions of pure states has also a bearing on joint extensions of states as considered in the previous section~\cite{wer}. Indeed, suppose that $\rho_{12}$ and $\rho_{23}$ are pure and agree on the middle system, which is easily feasible, then a joint extension $\rho_{123}$ can only exist for $\rho_{12}$ and $\rho_{23}$ pure product states. Therefore a generic pure two-party state with inner shift-invariance as in~(\ref{inv}) cannot be extended.

Suppose that density matrices $\rho_{12}$ and $\sigma_{23}$ agree on the middle system and can be jointly extended. The set of extensions is still convex and compact and so we can still look for the maximal entropy extension. Finding this state is hard, however, because generally
\begin{equation*}
\bigl[ \rho_{12} \otimes \idty_3 \,,\, \idty_1 \otimes \sigma_{23} \bigr] \ne 0
\end{equation*}
or, equivalently, if $\rho_{12}$ and $\sigma_{23}$ are equilibrium states corresponding to Hamiltonians $h_{12}$ and $h_{23}$
\begin{equation*}
\tr_3 \exp\bigl( h_{12} + h_{23} \bigr) \not\approx \exp\bigl( h_{12} + h_2 \bigr).
\end{equation*}
Moreover, the maximal entropy extension will not saturate the strong sub-additivity.

Actually, a nice characterisation of equality in strong sub-additivity for a state $\rho_{123}$ on a space $\c H_1 \otimes \c H_2 \otimes \c H_3$ in terms of decompositions of the middle space has been obtained in the paper by Hayden et {al.~\cite{hay}}. The necessary and sufficient condition is that the middle Hilbert space $\c H_2$ decomposes as
\begin{equation*}
\c H_2 = \loplus_\alpha \c H_{\text{left}}^\alpha \otimes \c H_{\text{right}}^\alpha \enskip\text{and}\enskip \rho_{123} = \loplus_\alpha \lambda_\alpha\, \rho_{12}^\alpha \otimes \rho_{23}^\alpha
\end{equation*}
with $\{\lambda_\alpha\}$ convex weights.

\subsection*{A Qubit Example with SU(2)-symmetry}

An example of the limitations imposed on quantum state extensions can be worked out for qubits with a SU(2)-symmetry. In order to impose SU(2)-symmetry on single qubit observables we use the adjoint representation of SU(2)
\begin{equation*}
\r{Ad}(U): A \mapsto U\,A\,U^*,\enskip U \in \r{SU(2)},\enskip A \in \c M_2.
\end{equation*}
This is a reducible representation that decomposes into a spin 0 and a spin 1 irrep:
\begin{equation*}
\c M_2 = \Cx \idty \oplus \Cx \bit\sigma.
\end{equation*}
The only SU(2)-invariant state on $\c M_2$ is the uniform state
\begin{equation*}
\rho = \tfrac{1}{2}\, \idty.
\end{equation*}

For 2 qubits $\r{Ad}(U \otimes U)$ decomposes into 2 spin 0, 3 spin 1 and 1 spin 2 irrep. There exists now a one-parameter family of SU(2)-invariant states
\begin{equation*}
\rho = \tfrac{1}{3}\,(1-\lambda) (\idty - \p) + \lambda\, \p,\enskip 0 \le \lambda \le 1.
\end{equation*}
Here $\p$ is the projector on the singlet vector $\tfrac{1}{\sqrt 2}\, (|10\> - |01\>)$ in $\Cx^2 \otimes \Cx^2$. This projector commutes with every unitary of the form $U \otimes U$ and every two-qubit observable that is SU(2)-invariant is a linear combination of $\p$ and $\idty$. Clearly, SU(2)-invariant two-qubit states satisfy
\begin{equation*}
0 \le \< \p \> = \lambda \le 1.
\end{equation*}
The two-qubit state $\rho$ is separable for $0 \leq \lambda \leq \frac{1}{2}$ and entangled for $\frac{1}{2} < \lambda \leq 1$. Hence the expectation value of this projector for a certain process tells us how much bipartite entanglement between two neighbouring spins is attainable.

For 3 qubits the SU(2)-invariant states can still easily be determined but things become more complicated with increasing number of parties. Let
\begin{equation*}
\begin{split}
&\p_1 = \p \otimes \idty \enskip\text{and}\enskip \p_2 = \idty \otimes \p \enskip\text{and put} \\
&\q = \tfrac{4}{3}\, \bigl( \p_1 + \p_2 - \p_1 \p_2 - \p_2 \p_1 \bigr).
\end{split}
\end{equation*}
The algebra of three-qubit observables that are SU(2)-invariant is not Abelian. It can be decomposed into a direct sum of $\Cx$ and $\c M_2$ where $\Cx$ is identified with $\Cx \q$ and $\c M_2$ with the algebra generated by $\p_1$ and $\p_2$, not including $\idty$. An SU(2)-invariant three-qubit state is of the form
\begin{equation*}
\rho = \tfrac{1}{4}\, (1-\lambda) (\idty - \q) + \lambda \bigl( a\, \p_1 + b\, \p_2 + c\, \p_1 \p_2 + \overline c\, \p_2 \p_1 \bigr)
\end{equation*}
with 
\begin{equation*}
0 \le \lambda \le 1,\enskip a,b \in \Rl,\enskip c \in \Cx,\enskip 2a + 2b + \Re\g e(c) = 1, \enskip\text{and}\enskip \abs c^2 \le 4ab.
\end{equation*}
If we look for a SU(2)-invariant three-party state with partial shift-invariance, then we find the following constraint on the expectation of $\p$
\begin{equation}
0 \le \< \p_1 \> = \< \p_2 \>\le 3/4.
\label{p:3}
\end{equation}

SU(2) and shift-invariant states on more parties will satisfy stronger upper bounds on the expectations of $\p$, see~(\ref{p:3}). Ultimately, if we look for a shift-invariant extension on the full half-chain then, using the Bethe {Ansatz~\cite{bet}}, one can show that
\begin{equation}
0 \le \< \p \> \le \log 2 \approx 0.69.
\label{p:inf}
\end{equation}

We may look for the largest expectation value of $\p$ that can be obtained within classes of shift-invariant states that can easily be handled. Consider as a first example point-wise limits of shift-invariant product states. Such states are actually invariant under arbitrary finite permutations of sites on the half-chain and are usually called exchangeable. Using the Bloch parametrization~(\ref{bloch}) and
\begin{equation*}
\p = \tfrac{1}{4} (\idty - \bit\sigma_1 \cdot \bit\sigma_2) \enskip\text{with}\enskip \bit\sigma_1 = \bit\sigma \otimes \idty \text{ and } \bit\sigma_2 = \idty \otimes \bit\sigma
\end{equation*}
we have to maximize
\begin{equation*}
\b x \in \Rl^3 \mapsto \tfrac{1}{16}\, \tr \bigl[ (\idty + \b x \cdot \bit\sigma_1) (\idty + \b x \cdot \bit\sigma_2) (\idty - \bit\sigma_1 \cdot \bit\sigma_2) \bigr]
\end{equation*}
subject to the constraint $\norm{\b x} \le 1$. It is easily seen that the maximum is reached for $\b x = 0$ for which value $\<\p\> = \tfrac{1}{4}$. Hence
\begin{equation}
\<\p\> \le \tfrac{1}{4} \enskip\text{for exchangeable states.}
\label{exc}
\end{equation}

The largest expectation for $\p$ that can be reached within the class of product states is
\begin{equation*}
\max\Bigl( \tfrac{1}{16}\, \tr \bigl[ (\idty + \b x_1 \cdot \bit\sigma_1) (\idty + \b x_2 \cdot \bit\sigma_2) (\idty - \bit\sigma_1 \cdot \bit\sigma_2) \bigr] \Bigr)
\end{equation*}
subject to the constraint $\norm{\b x_1}, \norm{\b x_2} \le 1$. The maximum $\tfrac{1}{2}$ is attained for $\b x_1 = - \b x_2 = \b x$ where $\b x \in \Rl^3$ is an arbitrary vector of length 1. Therefore
\begin{equation*}
\<\p\> \le \tfrac{1}{2} \enskip\text{for separable states.}
\end{equation*}
Moreover, this maximum is attained for shift-invariant separable states that are equal weight mixtures of period-2 product states
\begin{equation}
\tfrac{1}{2}\, |e_0\>\<e_0| \otimes |e_1\>\<e_1| \otimes |e_0\>\<e_0| \otimes \cdots + \tfrac{1}{2}\, |e_1\>\<e_1| \otimes |e_0\>\<e_0| \otimes |e_1\>\<e_1| \otimes \cdots,
\label{neel}
\end{equation}
where $\{e_0, e_1\}$ is any orthonormal basis in $\Cx^2$. Hence
\begin{equation*}
\<\p\> \le \tfrac{1}{2} \enskip\text{for shift-invariant separable states}
\end{equation*}
is an optimal upper bound. States of the form~(\ref{neel}) are extreme shift-invariant states which allow a convex decomposition in clustering period-2 states. This is called N\'eel order of period 2. The value $\frac{1}{2}$ for shift-invariant separable states is still not close to the maximum value of $\log 2$. One can get closer by constructing more general quantum processes.

\section{Constructing Processes}

We now turn to the construction of classical and quantum processes using as initial data a unity preserving CP map $\Gamma: \c M_d \to \c M_d$ with invariant state $\rho$. In the classical case this reduces to a stochastic matrix $T$ with invariant measure $\bit\mu$. The construction is based on finitely correlated {states~\cite{fan}}, also called matrix product states. These are processes for which the correlations across any link can be modelled by a finite dimensional vector space. These states are more general than the ones we have considered until now and are easily constructible in the thermodynamic limit, unlike the Bethe Ansatz states. The finitely correlated states where e.g.\ used in the lectures by J.~Eisert under the form of tensor networks. Actually pure states have been considered there as these lectures were focusing on ground states. We present here a different version that is adapted to mixed states.

The starting point is a unity preserving completely positive (UPCP) map
\begin{equation*}
\Lambda: \c M_d \otimes \c M_d \to \c M_d
\end{equation*} 
that is compatible with the given $\Gamma$ in the following sense
\begin{equation}
\Lambda(A \otimes \idty) = \Lambda(\idty \otimes A) = \Gamma(A),\enskip A \in \c M_d.
\label{comp}
\end{equation}
A process $\bit\omega$ is now generated by repeatedly contracting the local observables on the half-chain. Consider a sequence of UPCP maps
\begin{equation}
\begin{split}
&\Lambda^{(0)} := \Lambda: \c M_d \otimes \c M_d \to \c M_d \\
&\Lambda^{(1)} := \Lambda \circ (\Lambda \otimes \id): \c M_d \otimes \bigl( \c M_d \otimes \c M_d \bigr) \to \c M_d \\
&\vdots \\
&\Lambda^{(n)} := \Lambda \circ \bigl( \Lambda^{(n-1)} \otimes \id \bigr): \c M_d \otimes \bigl( \underbrace{\c M_d \otimes \cdots \otimes \c M_d}_{\text{$(n+1)$ times}} \bigr) \to \c M_d.
\end{split}
\label{fcs0}
\end{equation}
The expectation of a local observable $A_n \in \otimes_0^n \c M_d$ is then computed as
\begin{equation}
\bit\omega\bigl( A_n \bigr) := \tr \Bigl\{ \rho\, \Lambda^{(n)}(\idty \otimes A_n) \Bigr\}.
\label{fcs}
\end{equation}

To define a stationary process, (\ref{fcs}) must satisfy a number of requirements. The definition should be consistent in the first place, namely $\bit\omega(A_n \otimes \idty) = \bit\omega(A_n)$. This follows from the compatibility~(\ref{comp}) and the invariance of $\rho$:
\begin{align*}
\bit\omega(A_n \otimes \idty) 
&= \tr \Bigl\{ \rho\, \Lambda^{(n+1)}(\idty \otimes A_n \otimes \idty) \Bigr\} \\
&= \tr \Bigl\{ \rho\, \Bigl( \Lambda \circ \bigl( \Lambda^{(n)} \otimes \id \bigr)\Bigr)(\idty \otimes A_n \otimes \idty) \Bigr\} \\
&= \tr \Bigl\{ \rho\, \Lambda \Bigl( \Lambda^{(n)} (\idty \otimes A_n) \otimes \idty \Bigr) \Bigr\} \\
&= \tr \Bigl\{ \rho\, \Gamma\Bigl( \Lambda^{(n)} (\idty \otimes A_n) \Bigr) \Bigr\} \\
&= \tr \Bigl\{ \rho\, \Lambda^{(n)} (\idty \otimes A_n) \Bigr\} \\
&= \bit\omega(A_n).
\end{align*}
Next, we need positivity. This follows immediately from the complete positivity of $\Lambda$. The compatibility condition implies that $\Lambda$ maps the identity on $\Cx^d \otimes \Cx^d$ to the identity on $\Cx^d$. This implies the normalization and stationarity of $\bit{\omega}$.

It is important to observe that compatibility~(\ref{comp}) imposes a severe restriction on $\Gamma$. Not every UPCP transformation $\Gamma$ of $\c M_d$ admits a compatible extension. Moreover, the compatible extensions of $\Gamma$, whenever such extensions exist, form a compact and convex set and one may wonder about particular extensions. We now turn to some classes of examples.

\subsection{Hidden Markov Processes}

A classical observable, i.e., a $\Rl$-valued function $f$ on configuration space $\Omega = \{1,2,\ldots,d\}$ is naturally tabulated into a vector $\b f = \bigl( f(1), f(2), \ldots, f(d) \bigr)^{\s T} \in \Rl^d$ and identified with a diagonal matrix in $\c M_d$ through the map
\begin{equation*}
\r{dia}(\b f) = \sum_\epsilon f(\epsilon)\, |\epsilon\>\<\epsilon|.
\end{equation*}
The relation between a (completely) positive transformation $\Gamma$ of $\c M_d$ and a stochastic $d \times d$ matrix is then
\begin{equation*}
\Gamma\bigl( \r{dia}(\b f) \bigr) = \r{dia}(T\, \b f).
\end{equation*}
This allows to rewrite the compatibility equation~(\ref{comp}): a stochastic matrix $S: \Rl^d \otimes \Rl^d \to \Rl^d$ is compatible with a stochastic matrix $T: \Rl^d \to \Rl^d$ if
\begin{equation*}
\sum_{\epsilon_2} S_{\varphi, (\epsilon, \epsilon_2)} = \sum_{\epsilon_1} S_{\varphi, (\epsilon_1, \epsilon)} = T_{\varphi,\epsilon},\enskip \forall\ \varphi, \epsilon.
\end{equation*}
Let us introduce $d$ square matrices of dimension $d$ with non-negative entries
\begin{equation*}
E(\epsilon)_{\varphi,\eta} = S_{\varphi, (\eta, \epsilon)}.
\end{equation*}
The process generated by $S$ is then seen to be
\begin{equation*}
\bit\omega(\epsilon_0, \epsilon_1, \ldots, \epsilon_n) = \< \bit{\mu} \,,\, E(\epsilon_0) E(\epsilon_1) \cdots E(\epsilon_n) \b 1\>,
\end{equation*}
where $\b 1 \in \Rl^d$ has all its entries equal to one and $\bit{\mu}$ is the invariant probability vector for $T$.

A stochastic matrix $T$ always allows the extension
\begin{equation*}
S_{\varphi, (\eta, \epsilon)} = \delta_{\eta, \epsilon}\, T_{\varphi, \epsilon}.
\end{equation*}
The corresponding process is the usual Markov process~(\ref{mar:pro2}). More general extensions $\bit\omega$ are hidden Markov processes: there exists a larger configuration space $\Omega_1$, a function $F: \Omega_1 \to \Omega$ and a Markov process $\bit\omega_1$ on $\Omega_1$ such that
\begin{equation*}
\bit\omega(\epsilon_0, \epsilon_1, \ldots, \epsilon_n) = \sum_{F(\varphi_j) = \epsilon_j} \bit\omega_1(\varphi_0, \varphi_1, \ldots, \varphi_n).
\end{equation*} 
The entropy of hidden Markov processes can be computed using a method due to {Blackwell~\cite{bla,fns}}. The starting point is the asymptotic entropy production formula~(\ref{ent:pro}). The construction of the process, adding one point at a time, see~(\ref{fcs0}) and (\ref{fcs}), defines a dynamical system on the length-$d$ probability vectors. The entropy of the process is then obtained as an average over entropies of probability vectors with respect to the invariant measure of the dynamical system. Numerical evidence suggests that the Markov extension has the smallest entropy amongst all.

\subsection{Qubits with SU(2)-invariance cont.}

In order to have manifest SU(2)-invariance of the process we impose SU(2)-covariance both on the CP transformation of $\c M_2$ and on its compatible extensions from $\c M_2 \otimes \c M_2$ to $\c M_2$. Let $\c G \mapsto U_g$ be a unitary representation of a group $\c G$ on a Hilbert space $\c H$. The adjoint representation lifts it to a representation of $\c G$ on the bounded linear transformations $\c B(\c H)$ of $\c H$:
\begin{equation*}
\r{Ad}(U_g)(A) = U_g A\, U_g^*,\enskip g \in \c G,\ A \in \c B(\c H).
\end{equation*}
Given two unitary representations $U^{(1)}$ and $U^{(2)}$ of $\c G$ on $\c H_1$ and $\c H_2$ a map $\Gamma: \c B(\c H_1) \to \c B(\c H_2)$ is covariant if
\begin{equation}
\Gamma \circ \r{Ad}(U^{(1)}) = \r{Ad}(U^{(2)}) \circ \Gamma
\label{cov}
\end{equation}

The Choi-Jamio\l kowski encoding of a linear map $\Gamma: \c M_{d_1} \to \c M_{d_2}$ is very convenient for handling complete positivity
\begin{equation*}
\s C(\Gamma) := \sum_{i,j} |i\>\<j| \otimes \Gamma(|i\>\<j|),
\end{equation*}
$\Gamma$ is completely positive if and only if $\s C(\Gamma)$ is positive semi-definite. The encoding depends on the chosen basis through the matrix units $|i\>\<j|$ but only up to unitary equivalence as
\begin{equation*}
\begin{split}
&\s C(\Gamma \circ \r{Ad}(U)) = \r{Ad}(U^{\s T} \otimes \idty) \circ \s C(\Gamma) \enskip\text{and} \\
&\s C(\r{Ad}(U) \circ \Gamma) = \r{Ad}(\idty \otimes U) \circ \s C(\Gamma).
\end{split}
\end{equation*}
The covariance condition~(\ref{cov}) for $\Gamma: \c M_{d_1} \to \c M_{d_2}$ translates for its Choi-Jamio\l kowski encoding into
\begin{equation*}
\bigl[ \overline U^{(1)}_g \otimes U^{(2)}_g \,,\, \s C(\Gamma) \bigr] = 0,\enskip g \in \c G.
\end{equation*}
Here $\overline A$ is the complex conjugate of the matrix $A$. For SU(2) there is an additional simplification because the conjugate of SU(2) is unitarily equivalent to SU(2).

It turns out that there is a one-parameter family of SU(2)-covariant UPCP transformations of $\c M_2$
\begin{equation*}
\Gamma(\bit\sigma) = \mu\, \bit\sigma,\enskip -\tfrac{1}{3} \le \mu \le 1.
\end{equation*}
The SU(2)-covariant UPCP maps $\Lambda: \c M_2 \otimes \c M_2 \to \c M_2$ compatible with $\Gamma$ are parametrized by three real parameters
\begin{equation}
\begin{split}
&\Lambda(\bit\sigma_1 \cdot \bit\sigma_2) = \alpha\, \idty, \\
&\Lambda(\bit\sigma_1) = \Lambda(\bit\sigma_2) = \Gamma(\bit\sigma) = \mu\, \bit\sigma, \enskip\text{and} \\
&\Lambda(\bit\sigma_1 \times \bit\sigma_2) = \eta\, \bit\sigma.
\end{split}
\label{cp:qubit}
\end{equation}
Complete positivity imposes constraints on $\alpha$, $\mu$, and $\eta$
\begin{equation}
\abs{6 \mu - \alpha} \le 3 \enskip\text{and}\enskip 3 - 2 \alpha -\alpha^2 + 12 \mu -12 \alpha \mu - 9 \eta^2 \ge 0.
\label{choi:qubit}
\end{equation}
These conditions can be obtained by imposing positivity on the Choi matrix of $\Lambda$. The allowed region is a piece of a cone in $\Rl^3$. We then compute the expectation of $\p$
\begin{align*}
\<\p\>
&= \tfrac{1}{4} - \tfrac{1}{4}\, \< \bit\sigma_1 \cdot \bit\sigma_2\> = \tfrac{1}{4} - \tfrac{1}{8}\, \tr \sum_\gamma \Lambda\bigl( \sigma^\gamma \otimes \Lambda(\sigma^\gamma \otimes \idty) \bigr) \\
&= \tfrac{1}{4} - \tfrac{1}{8}\, \mu \tr \Lambda(\bit\sigma_1 \cdot \bit\sigma_2) = \tfrac{1}{4}\, (1 - \alpha\mu).
\end{align*}
The maximum in the allowed parameter region is attained for $\alpha = - \tfrac{3}{2}$ and $\mu = \tfrac{1}{4}$ and is independent of $\eta$. Therefore
\begin{equation}
\<\p\> \le \tfrac{11}{32}
\end{equation}
for $\<\ \>$ a stationary and SU(2)-invariant process as in~(\ref{fcs0}). This should be compared with~(\ref{exc}).

In passing from exchangeable to shift-invariant separable states we actually allowed product states of period 2. This can also be applied to processes of the type~(\ref{cp:qubit}). Considering $\bit\sigma_1 \cdot \bit\sigma_2$ as the contribution to the energy of two neighbouring spins, a minimal value of $\<\bit\sigma_1 \cdot \bit\sigma_2\>$ corresponds to a maximal value of $\<\p\>$ and this is expected to happen for spins as anti-parallel as possible. Therefore the second requirement in~(\ref{cp:qubit}) is inappropriate and we should consider general SU(2)-covariant maps $\Lambda: \c M_2 \otimes \c M_2 \to \c M_2$. These are determined by four real parameters
\begin{equation}
\begin{split}
&\Lambda(\bit\sigma_1 \cdot \bit\sigma_2) = \alpha\, \idty, \\
&\Lambda(\bit\sigma_1) =  \mu\, \bit\sigma,\enskip \Lambda(\bit\sigma_2) = \nu\, \bit\sigma, \enskip\text{and} \\
&\Lambda(\bit\sigma_1 \times \bit\sigma_2) = \eta\, \bit\sigma.
\end{split}
\label{cp2:qubit}
\end{equation}
Complete positivity imposes the constraints
\begin{equation}
\begin{split}
&\abs{3 \mu + 3 \nu - \alpha} \le 3 \enskip\text{and} \\ 
&3 - 2 \alpha -\alpha^2 + 6 (1 - \alpha) (\mu + \nu) - 9 (\mu - \nu)^2 - 9 \eta^2 \ge 0.
\end{split}
\label{choi2:qubit}
\end{equation}
We now introduce two SU(2)-covariant maps $\Lambda_i: \c M_2 \otimes \c M_2 \to \c M_2$ as in~(\ref{cp2:qubit}) with defining parameters $(\alpha_i, \mu_i, \nu_i, \eta_i)$. The expectation of $\p$ in the equal weight average of these period-2 processes is given by
\begin{equation*}
\<\p\> = \tfrac{1}{4} - \tfrac{1}{8}\, (\alpha_2 \mu_1 + \alpha_1 \nu_2).
\end{equation*}
Maximizing this in the allowed parameter region yields
\begin{equation*}
\<\p\> = \tfrac{5}{8} = 0.625
\end{equation*}
which is within 10\% of the optimal bound and well within the entangled shift-invariant states.

\subsection{Davies Maps}

An interesting and physically relevant class of channels are the Davies maps, they arise in the reduced dynamical description of a system with a discrete level structure weakly coupled to a thermal {bath~\cite{ali}}. The level structure of the small system is preserved in the  sense that such a map $\Gamma$ is parametrized by a stochastic map $T$ and a decoherence matrix $D$. The matrix $T$ describes the stochastic evolution of the diagonal elements while $D$ gives the damping of the off-diagonal terms. Assuming that the system Hamiltonian is diagonal in the canonical basis
\begin{equation}
\Gamma\bigl( \r{dia}(\varphi) \bigr) = \r{dia}(T\,\varphi) \enskip\text{and}\enskip \Gamma(e_{ij}) = D_{ij} e_{ij},\enskip i \neq j.
\label{detbal}
\end{equation}
Here, $e_{ij} = |e_i\> \<e_j|$. Moreover, $T$ is detailed balance and $D$ is real symmetric. Detailed balance means that $T$ is Hermitian for the stationary measure $\bit\mu$ that is interpreted as the Gibbs state of the system
\begin{equation*}
\bit\mu(f\, T(g)) = \bit\mu(T(f)\,g),\enskip \text{$f$ and $g$ real-valued}.
\end{equation*}
This condition is equivalent with micro-reversibility: the occupation rate of level $i$ times the jump probability from $i$ to $j$ is equal to the occupation rate of $j$ times the jump probability from $j$ to $i$. Another equivalent condition is
\begin{equation*}
T_{ij} T_{jk} T_{ki} = T_{ik} T_{kj} T_{ji}\enskip \text{for all choices of $i,j,k$}.
\end{equation*}

Complete positivity additionally imposes that
\begin{equation}
\begin{pmatrix}
T_{11} &D_{12} &\cdots &D_{1d} \\
D_{21} &T_{22} &\cdots &D_{2d} \\
\vdots &\vdots &\ddots &\vdots \\
D_{d1} &D_{d2} &\cdots &T_{dd}
\end{pmatrix}
\enskip \text{be positive semi-definite.}
\label{davies:cp}
\end{equation}
The action of $\Gamma$ is quite clear: decay of the off-diagonal elements and birth and death process for the diagonals. The relations between the different rates are encoded in the positivity condition~(\ref{davies:cp}). E.g., one can readily check that the decay rate of the off-diagonals cannot be less than half the rate of convergence to equilibrium for the diagonal process.

For Davies maps one could expect the standard basis vectors to be minimizers of output entropy but this is not generally true. It has been shown~\cite{rog} that already for a single qubit a true superposition of ground and excited state is the minimizer in a regime where the map is close to the identity map and so truly quantum. High powers of a Davies map converge to the projector on the equilibrium state which is entanglement breaking. In this regime the minimizer for output entropy is the state corresponding to the row of minimal entropy in $T$.

The construction of a process as in~(\ref{fcs0}) requires a Davies map rather closer to the projector on the equilibrium state than to the identity. For a single qubit
\begin{equation*}
T = \begin{pmatrix} 1-a &a \\b &1-b \end{pmatrix}, \enskip\text{with}\enskip 0 \le b \le a \le 1
\end{equation*}
and with $d$ the damping factor of the off-diagonal element one checks that it generates a process if
\begin{equation*}
d^2 \le \tfrac{1}{2}\, (1-a)(1-b).
\end{equation*}

\subsection{Free Fermionic Processes}

For both Bosons and Fermions there exists a notion of Gaussian states and maps~\cite{ali2,bra} that are considerably simpler to handle than general ones. The names free, quasi-free, quadratic, linear, and determinantal are also used. Moreover, these states and maps are good approximations whenever the statistics dominates over the interactions. A considerable benefit is also the scaling behaviour: the dimension of the free objects grows linearly in the number of particles instead of exponentially. We shall here only describe the defining free objects without connecting them to the true observables, states, and maps of a many particle system. This yields a kind of meta-description. Moreover, we restrict ourselves to Fermions.

The observables of Fermions with mode space $\c H$ are the trace-class operators $\c I_1(\c H)$ on $\c H$. Apart from their linear structure commutators are also useful. The mode space $\c H_{12}$ of a bipartite system is just the direct sum of the mode spaces $\c H_1$ and $\c H_2$ of the corresponding subsystems and observables of system 1 are extended by putting
\begin{equation*}
A \in \c I_1(\c H_1) \mapsto A \oplus 0 \in \c I_1(\c H_{12}).
\end{equation*}

The symbols play the role of states, they are operators $Q$ on $\c H$ such that $0 \le Q \le \idty$. The expectation of an observable $A$ is just $\tr Q\,A$. Mixtures of symbols are constrained by the following requirement: let $0 < \lambda < 1$ then the mixture of $Q_1 \ne Q_2$ with weights $\lambda$ and $1-\lambda$ can be formed if and only if $Q_1 - Q_2$ is of rank 1 and it yields the symbol $\lambda Q_1 + (1-\lambda) Q_2$. It then follows that a symbol is pure if and only if it is an orthogonal projector, possibly 0. Given a composite system $\c H_{12} = \c H_1 \oplus \c H_2$ and two symbols $Q_1$ and $Q_2$, the product state has symbol $Q_1 \oplus Q_2$. In this context, a separable state is just block diagonal in the mode space decomposition. We shall also need the von~Neumann entropy of a symbol on a finite dimensional space
\begin{equation}
\s H(Q) := - \tr \bigl[ Q \log Q + (\idty - Q) \log (\idty - Q) \bigr].
\label{ent:free}
\end{equation}
In particular, $\s H(Q) = 0$ if and only if $Q$ is pure.

A trace-preserving completely positive map from $\c H_1 \to \c H_2$ is a couple $(A, B)$ of linear maps where
\begin{equation}
A: \c H_1 \to \c H_2,\enskip B: \c H_1 \to \c H_1,\enskip \text{and}\enskip 0 \le B \le \idty - A^*A.
\label{qf:cp}
\end{equation}
The action on a symbol is given by
\begin{equation*}
Q \mapsto Q' = A^*Q\,A + B.
\end{equation*}
Observe that such maps are compatible with the notion of convex mixture of above because there is only a single Kraus-like operator appearing in~(\ref{qf:cp}). Composition of free CP maps is given by a semi-direct product
\begin{equation*}
(A, B) \circ (A', B') = (A\,A', B' + (A')^* B\, A' ).
\end{equation*}

We now mimic within the Fermionic free context the construction of a process starting from a CP transformation $(A, B)$ of $\Cx^d$. Such a map can be extended to a compatible map $(C, D)$ from $\Cx^d \oplus \Cx^d \to \Cx^d$ if and only if $A^*A \le \min\bigl( \{\tfrac{1}{2}\, \idty, \idty - B\} \bigr)$ and the extensions are labelled by an $X: \Cx^d \to \Cx^d$
\begin{equation*}
C = \begin{pmatrix} A &A \end{pmatrix} \enskip\text{and}\enskip D = \begin{pmatrix} B &X \\ X^* &B \end{pmatrix}.
\end{equation*}
The symbol $Q$ invariant under $(A, B)$ is the solution of
\begin{equation*}
Q = A^*Q\,A + B.
\end{equation*}
It is not hard to see that the outcome of the construction is a symbol $Q_\infty$ on $\oplus_0^\infty \Cx^d$ which is a block Toeplitz matrix: the $d \times d$ matrix entries in $Q_\infty$ are constant along parallels to the main diagonal. Explicitly
\begin{equation}
\bigl( Q_\infty \bigr)_{i\,i} = Q \enskip\text{and}\enskip \bigl( Q_\infty \bigr)_{i\,i+n} = (A^*)^n (Q - B + X).
\label{qinf}
\end{equation}

The entropy can now be computed in two different ways, we can either compute the limiting average entropy as in~(\ref{av:ent}) or the asymptotic entropy production as in~(\ref{ent:pro}). The first method relies on an extension of Szeg\"o's theorem to block Toeplitz matrices. For the second we need either a much finer control on the spectra of principal sub-matrices of a block Toeplitz matrix, which appears to be hard, or we have to exploit the smoothness of the entropy function. We follow this last approach.

Let $T: [-\pi, \pi[ \to \c M_d$ be an $\c L^\infty$-function taking values in the Hermitian $d \times d$ matrices and put
\begin{equation*}
\hat T =
\begin{pmatrix}
\hat T(0) &\hat T(1) &\hat T(2) &\cdots \\
\hat T(-1) &\hat T(0) &\hat T(1) &\cdots \\
\hat T(-2) &\hat T(-1) &\hat T(0) &\cdots \\
\vdots &\vdots &\vdots &\ddots 
\end{pmatrix}
\end{equation*}
where $\hat T$ are the Fourier coefficients of $T$
\begin{equation*}
\hat T(k) := \tfrac{1}{2\pi}\, \int_{-\pi}^\pi \!d\theta\, T(\theta)\, \r e^{-2\pi ik\theta}.
\end{equation*}
An extension of the classical Szeg\"o theorem~\cite{mir} reads

\begin{theorem}[Szeg\"o]
For any continuous complex function $f$ on $\Rl$
\begin{equation}
\lim_{n \to \infty} \frac{1}{n}\, \tr f(P_n \hat T\, P_n) = \frac{1}{2\pi}\,\int_{-\pi}^\pi \!d\theta\, \tr f(T(\theta))
\label{szego}
\end{equation}
where $P_n$ projects onto the first $n$ terms in $\oplus_0^\infty \Cx^d$.
\end{theorem}

The theorem gives us information on the main asymptotic behaviour of the eigenvalues of principal sub-matrices of $\hat T$. Consider e.g., the case $d = 1$, then Szeg\"o's theorem can be rewritten as
\begin{equation*}
\lim_{n \to \infty} \frac{1}{n}\, \tr f(P_n \hat T\, P_n) = \int_{\Rl} \!\mu(dx)\, f(x)
\end{equation*}
where
\begin{equation*}
\mu(]-\infty,x]) = \frac{1}{2\pi}\, \int_{T(\theta) \le x} \!d\theta.
\end{equation*}
If we order the eigenvalue list $\{\lambda_{n,j}\}$ of $P_n \hat T\, P_n$, then $\{\lambda_{n,j}\}$ interlaces $\{\lambda_{n+1,j}\}$ and
\begin{equation*}
\r w^*\text{-}\lim \frac{1}{n}\, \sum_{j=1}^n \delta_{\lambda_{n,j}} = \mu.
\end{equation*}
A fine asymptotic control on the eigenvalues could be used to obtain the average~(\ref{szego}) as an asymptotic growth
\begin{equation*}
\lim_{n \to \infty} \bigl\{ \tr f(P_n \hat T\, P_n) - \tr f(P_{n-1} \hat T\, P_{n-1}) \bigr\}.
\end{equation*}
Numerical evidence, however, shows that the behaviour of eigenvalue spacings can become erratic when $T$ oscillates.

\begin{figure}
\begin{center}
\psfig{file=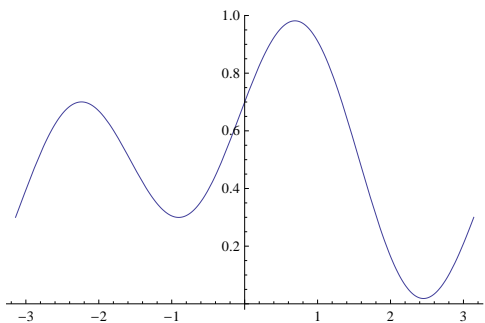,width=2.5in}
\end{center}
\caption{The function $T(\theta) = \tfrac{1}{2} + \tfrac{1}{5}\, \cos(\theta) + \tfrac{1}{3} \sin(2\theta)$}
\label{fig1}
\end{figure}

\begin{figure}
\begin{center}
\psfig{file=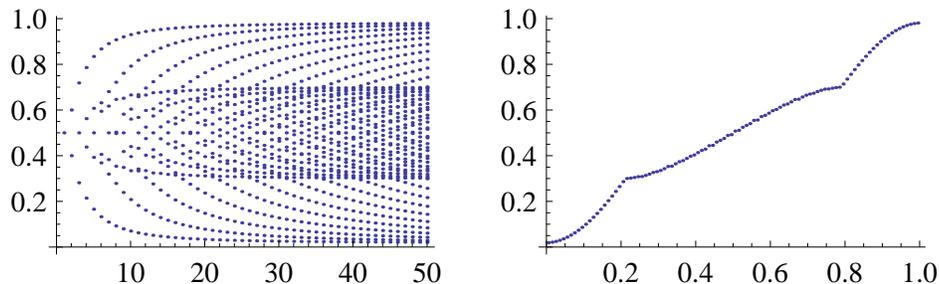,width=5in}
\end{center}
\caption{Eigenvalues of principal Toeplitz sub-matrices}
\label{fig2}
\end{figure}

In fig.~\ref{fig1} the function $T(\theta) = \tfrac{1}{2} + \tfrac{1}{5}\, \cos(\theta) + \tfrac{1}{3} \sin(2\theta)$ is plotted. In fig.~\ref{fig2} the eigenvalue lists of the first 50 principal sub-matrices are shown together with a plot of the eigenvalues of the $100 \times 100$ sub-matrix. This last plot approximates well the reordered function $T$.

Strong sub-additivity of the entropy guarantees the existence of the asymptotic entropy production but a much more general result can be proven by extending Szeg\"o's theorem to reasonably smooth functions. This result can then be applied to the computation of the entropy of our processes.

\begin{theorem}
For a block Toeplitz matrix $T$ and an absolutely continuous complex function $f$ on $\Rl$ 
\begin{equation*} 
\lim_{n \to \infty} \Bigl( \tr f(P_{n+1} \hat T P_{n+1}) - \tr f(P_n \hat T P_n) \Bigr) = \frac{1}{2\pi} \int_{-\pi}^\pi \!d\theta\, \tr f(T(\theta)).
\end{equation*}
\end{theorem}

As a corollary we get
\begin{equation*}
\s h = \tfrac{1}{2\pi}\, \int_{-\pi}^\pi \!d\theta\, \s H(Q_\infty(\theta))
\end{equation*}
with
\begin{equation*}
Q_\infty = Q + \frac{A^* \r e^{i\theta}}{\idty - A^* \r e^{i\theta}}\, (Q - B + X) + \text{h.c.}
\end{equation*}
and $\s H$ as in~(\ref{ent:free}).

\section{Conclusion}
The construction of quantum processes is a lot less straightforward than for their classical counterparts. The intricate nature of quantum correlations complicates even the seemingly simple task of finding extensions of overlapping states.

Processes that can nevertheless be easily constructed, like exchangeable or separable states, are not general enough to study interesting quantum behaviour. On the other hand, more general constructions like the Bethe Ansatz become difficult to handle as the size of the process increases.

The processes we have studied here, the finitely correlated states, lie somewhere in between the previous two classes. By construction, they are well-behaved as the length of the process grows. We have also seen by studying some concrete examples that such states do in fact exhibit interesting quantum characteristics.

\section*{Acknowledgements}

This work is partially funded by the Belgian Interuniversity Attraction Poles Programme P6/02.

\end{document}